\documentclass{elsart}     % for users with LaTeX2e

\input epsf

\begin{document}
\begin{frontmatter}
\title{Statistical physics and stromatolite growth: 
new perspectives on an ancient dilemma}
 
\author[ANUa]{M.T. Batchelor} 
\author[ANUb]{R.V. Burne} 
\author[UNSW]{B.I. Henry}
 and
\author[ANUa]{T. Slatyer}
\address[ANUa]{Department of Theoretical Physics, RSPSE\\ 
and Mathematical Sciences Institute,\\
The Australian National University, Canberra ACT 0200, Australia}
\address[ANUb]{Department of Geology, The Australian National University,\\
Canberra ACT 0200, Australia}
\address[UNSW]{Department of Applied Mathematics, School of Mathematics,\\
The University of New South Wales, Sydney NSW 2052, Australia}
 
\begin{abstract}
This paper outlines our recent attempts to model the growth and form of microbialites 
from the perspective of the statistical physics of evolving surfaces. 
Microbialites arise from the environmental interactions of microbial communities 
(microbial mats). 
The mats evolve over time to form internally laminated organosedimentary structures (stromatolites). 
Modern day stromatolites exist in only a few locations, whereas ancient 
stromatolitic microbialites were the only form of life for much of the Earth's history.
They existed in a wide variety of
growth forms, ranging from almost perfect cones to branched columnar structures.
The coniform structures are central to the heated debate on the oldest evidence 
of life. 
We proposed a biotic model which considers the relationship
between upward growth of a phototropic or phototactic biofilm and mineral 
accretion normal to the surface. 
These processes are sufficient 
to account for the growth and form of many ancient stromatolities. 
These include 
domical stromatolites and coniform structures with thickened apical zones typical 
of Conophyton. 
More angular coniform structures, similar to the stromatolites 
claimed as the oldest macroscopic evidence of life, form when the photic
effects dominate over mineral accretion.
\end{abstract}

%PACS: 87.10.+e; 89.20.-a; 91.90.+p}
%
%Keywords: Microbialites; Stromatolites; KPZ equation

\end{frontmatter}

\section{Introduction}

The Kardar, Parisi and Zhang (KPZ) equation \cite{KPZ} 
describing the temporal and spatial evolution of an interface
has found widespread applicability in a diverse range of phenomena \cite{BS}.
These include the description of evolving fronts in experiments on
paper burning, chemical electrodeposition and yeast colonies.
KPZ originally suggested similarities with geological stratification.
Now stromatolites are internally laminated biosedimentary 
structures \cite{LRG,Walter} produced
as a consequence of some of the environmental interactions of benthic
microbial communities (BMC) \cite{CR}.
These microbialites \cite{BM} have been described as Earth's default ecosystem \cite{Gould}, 
re-establishing themselves 
when other life-forms are extinguished by environmental stress \cite{Science}.  
Stromatolitic microbialites provide an almost continuous record of the subsequent 
evolution of environmental conditions on Earth.  
They flourish wherever and whenever conditions favour their mineralisation, 
and where they are neither out-competed by faster-growing higher organisms 
nor consumed by grazers.

Most importantly, 
stromatolitic microbialites  preserve the only macroscopic evidence of life 
prior to the appearance of macro-algae \cite{K}.  
The biogenicity of stromatolites older than 3,200 million years is 
unclear \cite{WBD,ARR,L,HGHT}.  
If they are indeed biotic, they are the oldest morphological evidence for life, 
now that the identification of 3,300 to 3,500 million year old microfossils \cite{SP} has been 
challenged \cite{B,G}.  
Central to this debate are the sharply peaked conical stromatolites 
described from the 3,450 million year old Warrawoona Group in  
Western Australia \cite{HGHT}.

We recently proposed a KPZ-like model for stromatolite morphogenesis that 
endorses a biotic origin for coniform stromatolites \cite{BBHJ}.
The model describes interaction between upward growth of a phototropic or 
phototactic microbial mat and 
mineral accretion normal to the surface of the mat.  
Domical structures are formed when mineral accretion dominates.  
When vertical growth dominates, coniform structures evolve that reproduce the 
features of Conophyton, a stromatolite that flourished in certain low-sedimentation 
environments for much of the Proterozoic \cite{BSMP}.  
Increasing the dominance of vertical growth produces sharply-peaked structures
similar to those described in the Warrawoona Group.
Here we consider this model in (2+1)-dimensions and give some preliminary results 
on the solutions to the model equation.

\section{The model}

The proposed model \cite{BBHJ} for stromatolite morphogenesis involves two processes 
only: 
(i) upward growth of a phototropic or phototactic BMC, 
(ii) mineral accretion normal to the surface. 
Here the function $h(x,y,t)$ represents the height of the profile above a horizontal baseplane which 
evolves in time $t$ according to the equation
\begin{equation} \frac{\partial h}{\partial t} = \frac{\lambda}{2} \left[ \left( 
\frac{\partial 
h}{\partial 
x}\right)^{2} + \left(\frac{\partial h}{\partial y}\right)^{2} \right] + \lambda 
+ v. \label{2dno_v} \end{equation}
We interpret $v$ as the average rate of vertical growth due to photic response of microbes and 
$\lambda$ as the average rate of surface-normal growth due to mineral accretion.

If only radially symmetric solutions are considered then the problem reduces to the
(1+1)-dimensional case after transforming to polar coordinates. 
Our approach to use the KPZ equation to describe stromatolites 
\cite{BBHJ,BBHW,BBHWd,BBHS} 
differs from others \cite{GR,GK} in that we interpret our model in purely biological terms.
We also disregard the usual KPZ noise term, thus allowing us to write down explict
solutions to the model equation.

\section{Preliminary results and discussion}

Although non-linear, equation (\ref{2dno_v}) can be solved with a change of 
variables using the 
method of characteristics \cite{BBHW} and prescribed initial profiles. 
The choice of initial profile is important. 
It is worth noting that cone-like profiles arise naturally in deformations of thin flat sheets \cite{LGLMW}. 
For simplicity, we consider here a radially symmetric cone centred on the origin with
base radius $R$ and initial height $h(0,0,0) = h_{0}$.\footnote{Full details will be given 
elsewhere \cite{BBHS}.}
Then the full solution to equation (\ref{2dno_v}) is given by
\begin{equation} h(x,y,t) = \left\{ \begin{array}{ll} h_{0} - \frac{r^{2}}{2 
\lambda t} + 
(v + 
\lambda)t & $if $ r \leq \frac{\lambda h_{0} t}{R}, \\ h_{0} \left(1 - 
\frac{r}{R}\right) + \frac{\lambda h_{0}^{2} t}{2 R^{2}} + (v + \lambda)t~~ & 
$if $ 
\frac{\lambda h_{0} t}{R} \leq r \leq R + \frac{\lambda h_{0} t}{2 R}, \\ 
(v + \lambda)t & $if $ r \geq R + \frac{\lambda h_{0} t}{2 R}, \end{array} 
\right. \end{equation}
where $r = \sqrt{x^{2} + y^{2}}$.
From the solution we see that the convex 
sharp protrusion at the tip of the initial cone gives rise to a smooth radially 
symmetric paraboloid centred on the origin, expanding outwards with time.  
The boundary of 
the cone's base represents a concave discontinuity in the spatial derivative, 
and gives rise to a shock which propagates outwards with time, effectively 
expanding the base of the cone.

In a similar way one can consider more complex cases, such as a paraboloid with
an elliptical base.
The paraboloid can be either convex or concave, the degree to which can be tuned 
by an adjustable parameter $\alpha$.
However, in all the cases we have studied, an isolated protrusion centred at the origin and 
evolving under equation (\ref{2dno_v}) approaches the smooth radially symmetric 
paraboloid solution,
\begin{equation} 
h(x,y,t) = h(0,0,0) - \frac{r^{2}}{2 \lambda t} + (v + 
\lambda) t. \label{longtermsol2} 
\end{equation}
For an initial profile consisting of a field of protrusions, the long-term 
solution will therefore consist of smooth radially symmetric paraboloids 
separated by shocks.
Indeed, a  feature of the KPZ equation is that the 
long-term solutions are radially symmetric,  
independent of the initial profiles.

\begin{figure}[p]
%\vskip 3mm
\centerline{
\epsfxsize=14cm
\epsfbox{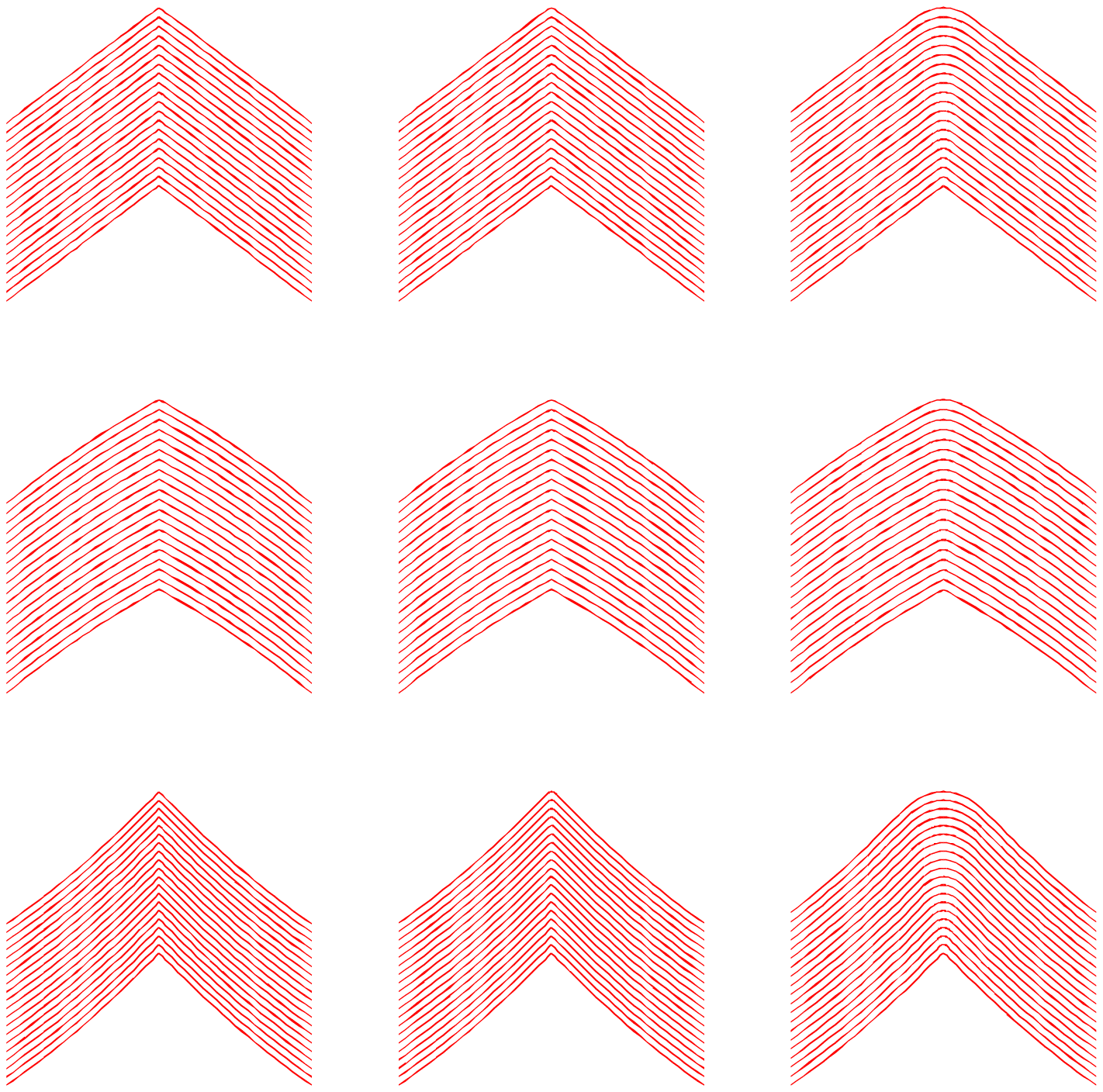}}
\caption{Equal time snapshots of the surface height profile for different values 
of the initial shape parameter $\alpha$ 
and the growth parameters $\lambda$ and $v$. First row: $\alpha = 20$, 
second row: $\alpha = 1.5$, third row: $\alpha = -1$. 
First column: $\lambda = 0.01, v = 1.99$, second column: $\lambda = 0.1, v = 
1.9$, third column: 
$\lambda = 1, v = 1$.}
\label{comparison}
\end{figure}

The parameter $\lambda$ controls the rate at which the shape of the paraboloid 
changes, as well as the rate of vertical growth (via the $(v + \lambda) t$ 
term). 
The parameter $v$, on the other hand, affects only the rate of vertical growth.  
The relative magnitudes of $\lambda$ and $v$ therefore determine the rate at 
which the paraboloid changes in shape relative to its rate of upward growth.  
For example, if $\lambda$ is very small compared to $v$, then even after a 
long period of time, as measured by the amount of upward growth, the paraboloid 
will still closely resemble its initial form.  
If $\lambda$ is comparable to or 
greater than $v$, the transition from the initial profile to a smooth 
radially symmetric paraboloid will take place over a much shorter period of 
vertical growth.  
These regimes are illustrated in the cross-sections shown in Figure~\ref{comparison}, which is to be compared
with Figure~1 of Ref.~\cite{BBHJ} obtained for the (1+1)-dimensional case.

The results provide possible explanations for variations in coniform stromatolite morphogenesis. 
Together with field evidence, they
support the interpretation that the vertical growth parameter $v$ represents photic response of 
the BMC rather than sediment deposition. 
If the converse were true, coniform stromatolites would only form under conditions of high 
sedimentation which is precisely contrary to field evidence. 
Indeed, while sediment deposition would tend towards the smoothing of surface irregularities, 
growth due to photic response would tend to accentuate them. 
Our model shows that a combination of vertical phototropic or phototactic microbial growth and 
surface-normal mineral accretion can produce coniform forms and structures analogous to those 
found in both Archaean and Proterozoic coniform stromatolites. 
For example, there is a striking similarity between the model forms shown in Figure~\ref{comparison} 
and the sharply-peaked coniform stromatolites in the Warrawoona Group \cite{HGHT}, 
thus supporting their 
biogenic origin and reinforcing the probability that photosynthetic microbes were components of 
Archaean BMCs.

\begin{figure}[t]
\vskip 3mm
\centerline{
\epsfxsize=2.8in
\epsfbox{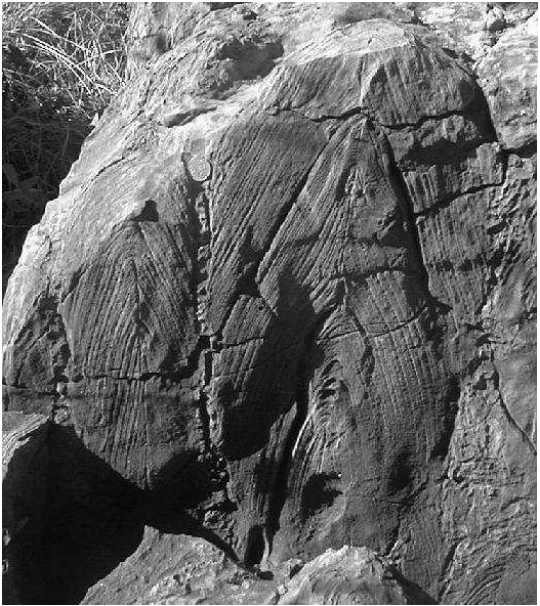}
\epsfxsize=2.8in
\epsfbox{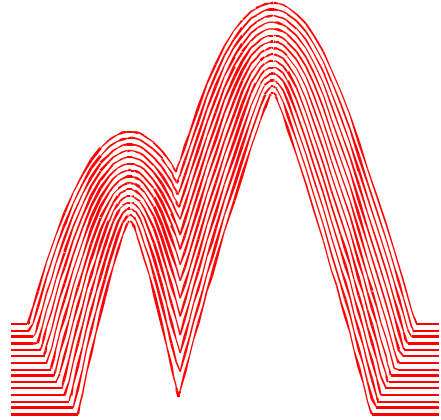}}
\caption{(left) 1,700 million year old Conophytons from the  Dungaminnie Formation, 
Northern Territory, Australia. Coin 28mm for scale.
(right) Longitudinal cross-section of a preliminary simulation that does not attempt to take
the complex paraboloid into account.}
\label{side}
\end{figure}

\begin{figure}[t]
\vskip -10mm
\centerline{
\epsfxsize=2.8in
\epsfbox{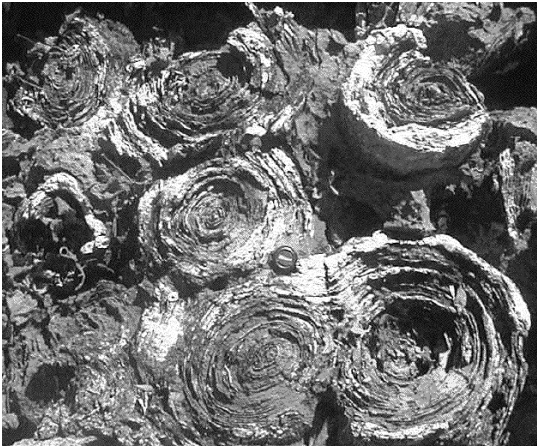}
\epsfxsize=2.8in
\epsfbox{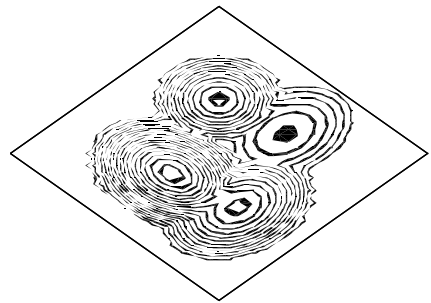}}
\caption{(left) 1,670 million year old Conophytons from the Macarthur
Group, Northern Territory, Australia. Lens cap 25mm for scale.
(right) Tansverse cross-section of a preliminary simulation.}
\label{top}
\end{figure}

The various cases modelled in Figure~\ref{comparison} can all be matched in Proterozoic Conophytons.
Figure~\ref{side} shows a comparison between Conophytons observed in the Northern Territory
of Australia and a 
longitudinal cross-section of a
preliminary simulation of two evolving contiguous cones.
Figure~\ref{top} shows a comparison between Conophytons and a transverse cross-section of 
an evolving paraboloid field.
Our aim here is to determine the environmental parameters responsible for the various
Conophyton growth forms.

We believe that our results also shed some light on why, after flourishing for much of the Proterozoic, Conophytons virtually disappeared in the Neoproterozoic \cite{KRS}. 
This demise has been linked to evolutionary changes in BMCs \cite{GK}, 
but since these would not have 
limited photic response, this seems untenable. 
Conophytons represent an effective growth strategy that is especially vulnerable to predation 
and competition \cite{JLR}.
Perhaps their demise is best explained in terms of the evolution of 
greater biological diversity in their quiet marine environments.

\ack
This work has been supported by the Australian National University and the Australian Research Council. 
MTB thanks Prof C.K. Hu for his kind hospitality at Academia Sinica, Taipei.

\clearpage

\end{document}